\documentclass{aa}

\usepackage{natbib}
\usepackage[switch]{lineno}
\usepackage{graphicx}
\usepackage{caption}
\usepackage{subcaption}
\usepackage{inputenc}
\usepackage{color}
\usepackage{amsmath,amsfonts,amssymb}
\usepackage{float}
\usepackage[colorlinks=true, citecolor=blue, linkcolor=blue, urlcolor=blue]{hyperref}
\usepackage[version=3]{mhchem}
\usepackage{xcolor}
\usepackage{booktabs}
\usepackage{multirow}

\newcommand{\mwm}{mW\,m$^{-2}$}
\newcommand{\keV}{keV}
\newcommand{\MeV}{MeV}

\begin{document}

\title{The precipitation of protons and electrons in Jupiter's auroral regions: a statistical comparison based on Juno/JEDI data}

\author{B. Benmahi\inst{1,2}
       \and N. Andre\inst{3,4}
       \and V. Hue\inst{1}
       \and Z.-Y. Liu\inst{3}
       \and G. Gronoff\inst{5}
       \and B. Mauk\inst{6}
       \and B. Bonfond\inst{2}
       \and T. Stallard\inst{7}
       \and M. Blanc\inst{8,3}
       \and M. Devinat\inst{3}
       \and M. Barthelemy\inst{9,10}
       }

\institute{Aix-Marseille Université, CNRS, CNES, Institut Origines, LAM, Marseille, France.\\ \email{bilal.benmahi@lam.fr}
\and Laboratory of Atmospheric and Planetary Physics, STAR Institute, University of Liège, Liège, Belgique
\and Institut de Recherche en Astrophysique et Planétologie (IRAP), CNES-CNRS-Université de Toulouse, Toulouse, France
\and Institut Supérieur de l'Aéronautique et de l'Espace (ISAE-SUPAERO), Université de Toulouse, Toulouse, France
\and NASA Langley Research Center, Hampton, Va, USA
\and Johns Hopkins University Applied Physics Laboratory, Laurel, MD USA
\and Northumbria University, Newcastle upon Tyne, NE1 8ST, UK
\and Weihai Institute for Interdisciplinary Research, Shandong University, Weihai, Chine
\and Univ. Grenoble Alpes, CNRS, IPAG, 38000 Grenoble, France
\and Univ. Grenoble Alpes, CSUG, 38000 Grenoble, France
}

\date{Received ??/??/??? ; Accepted ??/??/???}

\abstract
{Energetic electrons dominate the energy precipitated into Jupiter's auroral regions, but protons precipitate along the same field lines, with a contribution that remains poorly constrained.}
{We statistically compare the precipitating energy flux of protons and electrons in two of Jupiter's auroral sub-regions, the polar (PE) and main (ME) emission regions.}
{We use Juno/JEDI data from PJ3 to PJ66. The magnetic footprint of the spacecraft, computed with the JRM33 and CON2020 field models, is assigned to one of the two sub-regions, and the precipitating energy flux of each species is obtained by integrating over the loss cone the mean differential intensity measured within it.}
{In both sub-regions, the median proton energy flux is 52 to 223 times lower than that of the electrons, and the total precipitated proton power is less than 2\% of the electron power. Proton fluxes are broadly distributed across the sub-regions, with higher values in the ME than in the PE.}
{Protons play a secondary role in Jupiter's auroral energy budget, radiating an estimated 5 to 9 GW in the UV per sub-region, pending a dedicated proton transport modelling. Their Doppler signature in the red wing of the Lyman-$\alpha$ line, undetected by HST and difficult to access with Juno/UVS, is a key observable for the upcoming Juice and Europa Clipper missions.}

\keywords{Jupiter -- aurora -- magnetosphere -- particle precipitation -- energetic particles -- Juno}

\titlerunning{Proton and electron precipitation at Jupiter}
\authorrunning{B. Benmahi et al.}
\maketitle

\section{Introduction}
\label{sec:intro}
The interaction between Jupiter's magnetosphere and its atmosphere results in the precipitation of energetic charged particles along magnetic field lines toward the planet's polar regions \citep{cowley_origin_2001,hill_jovian_2001}. This precipitation generates the most intense auroral emissions in the solar system in the ultraviolet domain \citep{broadfoot_extreme_1979,clarke_observations_1980}, as well as infrared H$_3^+$ emissions \citep{drossart_detection_1989} and X-ray emissions from the polar cap \citep{cravens_implications_2003}. Beyond auroral emissions, precipitation also shapes the thermal structure of the upper atmosphere \citep{sinclair_jupiters_2017,sinclair_jupiters_2018}, influencing its dynamics \citep{Cavalie2021} and chemical composition \citep{cavalie_evidence_2023, Hue2024}. Jupiter's auroral regions are organized into three sub-regions, the main emission (ME), the polar emission (PE), and the outer emission (OE) (e.g.,~\citealp{grodent_brief_2015,benmahi_energy_2024,benmahi_auroral_2024}). Precipitation is dominated by electrons, whose energy distributions have been statistically characterized by \citet{salveter_jovian_2022} in the ME using the Jupiter Energetic Particle Detector Instrument aboard Juno (Juno/JEDI) data \citep{mauk_jupiter_2017}.

Energetic ions also precipitate along the same magnetic field lines as electrons. As early as the Voyager era, \citet{goertz_proton_1980} proposed that energetic proton precipitation could contribute to the Jovian aurora through Doppler-shifted Lyman-$\alpha$ emission. \citet{waite_role_1992} subsequently established, from high-resolution IUE Lyman-$\alpha$ profiles, that less than 10\% of Jupiter's UV auroral emissions could be attributed to this source. The first in situ detection of precipitating protons in the auroral regions was made during the first Juno perijove by \citet{haggerty_junojedi_2017}, who showed that the observed fluxes remain insufficient to explain the polar X-ray emissions, consistent with the modeling predictions of \citet{ozak_auroral_2013}.

Since then, event-based studies have shown that local acceleration and bidirectional transport along the field lines also contribute to the auroral proton population. Bidirectional fluxes have been reported at low altitude \citep{sulaiman_jupiters_2022}, together with outward beams near the ME \citep{szalay_proton_2021} and dispersive conics in the southern polar cap \citep{szalay_closed_2022}, while EMIC waves accelerate protons near the Io footprint \citep{clark_energetic_2020,clark_energetic_2023}. In the inner and middle magnetosphere, energetic protons exhibit pitch angle distributions predominantly perpendicular to the field, with nearly empty loss cones \citep{shen_energetic_2022}, which limits the rate at which magnetospheric protons reach the atmosphere. Each of these studies covers a single event or a small number of them. No systematic study has yet established a statistical map of the precipitating proton energy flux across the auroral sub-regions, nor quantitatively compared it to the electron flux over a large number of perijoves.

In this paper, we present the first statistical analysis of the precipitating proton energy flux in Jupiter's auroral regions using Juno/JEDI data from PJ3 to PJ66, and compare it to that of electrons to discuss the implications for the auroral energy budget and emissions.

\section{Data and method}
\label{sec:data}

\subsection{Instrument and data selection}
\label{sec:instrument}

This paper relies on level-3 (L3) calibrated data from JEDI \citep{mauk_jupiter_2017}, downloaded from the NASA Planetary Data System (PDS), covering the perijoves of Juno from PJ3 to PJ66. JEDI consists of three detector heads (JEDI-90, JEDI-180, and JEDI-270), each combining solid-state detectors (SSDs) with a time-of-flight (TOF) system, providing near-complete pitch angle coverage at a cadence of about 1\,s.

Electrons are measured from $\sim$30\,\keV{} to $\sim$1\,\MeV{} by single SSD measurements, while protons require positive identification through a coupled TOF-energy measurement, restricting the usable energy range to $\sim$50\,\keV{}--$\sim$1\,\MeV{}.

Before the analysis, the electron spectra are corrected for two instrumental effects. For all perijoves from PJ20 onward, the archived small pixel data below $\sim$50\,\keV{} are affected by a known processing error (B. Mauk, private communication), and this part of the spectrum is replaced by a power law fitted over [50\,\keV{}; 120\,\keV{}]. The Minimum Ionizing Bump (MIB), an artificial feature near $\sim$160\,\keV{} produced by highly energetic electrons passing through the SSDs \citep{mauk_jupiter_2017,mauk_juno_2026}, is removed by fitting the continuum between $\sim$100 and $\sim$280\,\keV{}. The electron dataset resulting from these corrections is the same as that used in a companion manuscript (Benmahi et al., in prep.), which addresses the north-south asymmetry and the up/down anisotropy of the electron precipitation, whereas the present paper uses it only as a reference for the proton fluxes.

We do not apply either of these corrections to the proton spectra. Ions within the nominal JEDI energy range do not penetrate the SSDs, so they cannot produce a genuine minimum ionizing feature. Only protons above $\sim$9\,\MeV{} would do so, at energies far beyond the range considered here (B. Mauk, private communication). In the intense radiation environment close to Jupiter, however, electrons can contaminate the ion channels through accidental coincidences in the TOF-energy measurement \citep{mauk_juno_2026}. Such coincidences are strongly suppressed in the TOF-energy product used here, which requires an additional solid-state detector signal, but they are not entirely absent in the harshest radiation regions. Since this contamination is instrumentally distinct from the electron MIB and cannot be identified at the spectrum level, these coincidences yielding apparently valid TOF-energy pairs, we retain the proton spectra as measured. The differential intensities used here are the L3 quantities distributed through the PDS, in which the TOF-energy detection efficiencies of the ion channels \citep{mauk_juno_2026} are already accounted for, so that no additional efficiency correction is applied in this work.

\subsection{Magnetic mapping and sub-region assignment}
\label{sec:mapping}

At each time step, Juno's magnetic footprint is computed by tracing the local magnetic field line down to Jupiter's atmosphere using the JRM33 internal field model \citep{connerney_new_2022} combined with the CON2020 current sheet contribution \citep{connerney_jovian_2020}, in System III coordinates. In both hemispheres, the ME boundaries are the contours derived for each perijove from Juno/UVS images \citep{groulard_dawn-dusk_2024}. The PE is the region poleward of the inner ME boundary, and the OE extends from the outer ME boundary to the mean magnetic footprint of Io, computed with the same field models.

Footprints are sampled on a $0.5^\circ \times 0.5^\circ$ grid for protons (versus $0.33^\circ \times 0.33^\circ$ for electrons) to compensate for their lower counting rate. The loss cone angle is computed identically for both species from the same field model combination. To limit contamination by penetrating radiation belt particles when Juno is far from the planet, even though the spacecraft's magnetic footprint remains connected to the auroral region at larger distances, we restrict the analysis to spectra measured within a planetocentric distance of $200\,000\,\text{km}$ ($\sim3\,R_J$).

The area of each sub-region is derived from its per-perijove boundaries and therefore varies from one perijove to the next. We use its median over PJ3--PJ66. In the northern hemisphere, the median areas are $(1.04\pm0.09)\times10^{9}$\,km$^2$ for the PE and $(3.61\pm0.47)\times10^{8}$\,km$^2$ for the ME. In the southern hemisphere, $(8.93\pm1.01)\times10^{8}$ and $(3.95\pm0.51)\times10^{8}$\,km$^2$ respectively, where the uncertainties are the perijove-to-perijove standard deviations. These relative dispersions, below $\sim$11\%, make the areas by far the best constrained term of the flux--area product used to estimate the precipitated powers.

Precipitating protons are detected in both sub-regions and in both hemispheres, with a final sample size comparable to that of electrons (7674 proton spectra against 7707 electron spectra). This near parity, however, does not reflect an equivalent detection efficiency per footprint: it partly results from the coarser $0.5^\circ$ sampling grid adopted for protons, chosen specifically to compensate for their markedly lower counting rate. Per footprint, protons remain detected far less often than electrons, a gap that does not stem from the geometric factors, which for ions are comparable to or larger than those of the small-pixel electron channels used here \citep[Table S6]{mauk_juno_2026}, but from the mandatory TOF-energy coincidence, whose efficiency peaks at about 0.3 near 90\,\keV{} and decreases at higher energies as $\mathrm{d}E/\mathrm{d}x$ \citep{mauk_juno_2026}.

\subsection{Computation of the precipitating flux}
\label{sec:flux}

Here $h_\downarrow(E)$ denotes the mean directional differential intensity within the downward loss cone, in particles\,cm$^{-2}$\,s$^{-1}$\,sr$^{-1}$\,keV$^{-1}$. The precipitating (downward) energy flux $\Phi_\downarrow$ (in \mwm{}) is
\begin{equation}
\Phi_\downarrow = \int_{E_\mathrm{min}}^{E_\mathrm{max}} E \cdot \pi\, h_\downarrow(E)\,\mathrm{d}E,
\label{eq:flux}
\end{equation}
where $[E_\mathrm{min}; E_\mathrm{max}]$ = [30\,\keV{}; 1\,\MeV{}] for electrons and [50\,\keV{}; 1\,\MeV{}] for protons; assuming isotropy within the loss cone, the factor $\pi$ results from the angular integration combined with the opening of the magnetic flux tube, and is independent of the loss cone half-angle. $\Phi_\downarrow$ is further mapped to a common reference altitude of $400\,\text{km}$, close to that of Jupiter's UV aurora \citep{bonfond_far-ultraviolet_2015}, assuming conservation of the energy flux along the flux tube, so that spectra acquired at different altitudes and in both hemispheres can be compared on a common basis. For each of the two sub-regions and each hemisphere, we build the statistical distributions of $\Phi_\downarrow$, normalized as a logarithmic density ($\mathrm{d}N/\mathrm{d}\log_{10}\Phi$), combining all spectra from PJ3 to PJ66.

The analysis presented here relies on JEDI alone, without the lower-energy populations measured by the Jovian Auroral Distributions Experiment (JADE, \citealp{McComas2017}). Two reasons motivate this choice. First, the integrand of Eq.~\ref{eq:flux} is weighted by the particle energy, so that the low-energy part of the spectrum contributes far less to the energy flux than it does to the particle density or to the shape of the distribution function. We have quantified this contribution for the electrons, using combined JADE and JEDI spectra available over a subset of these perijoves (PJ3--PJ35, about half the dataset used in this paper), which covers a comparable diversity of conditions to the full dataset. The median fraction of the electron energy flux carried below the JEDI threshold of $\sim$30\,\keV{} amounts to $\sim$6\% in both the PE and the ME. Second, combining the two instruments requires degrading the pitch angle resolution to their common sampling, which broadens the loss cone determination and substantially reduces the number of spectra that can be assigned to a given auroral sub-region. Using JEDI alone therefore preserves both the accuracy of the loss cone integration and the auroral coverage of the statistics, at a quantified and moderate cost on the energy flux itself.

\section{Results and Discussion}
\label{sec:results}

\subsection{Energy flux statistics}

\begin{figure*}[h!]
    \centering
    \includegraphics[width=18cm]{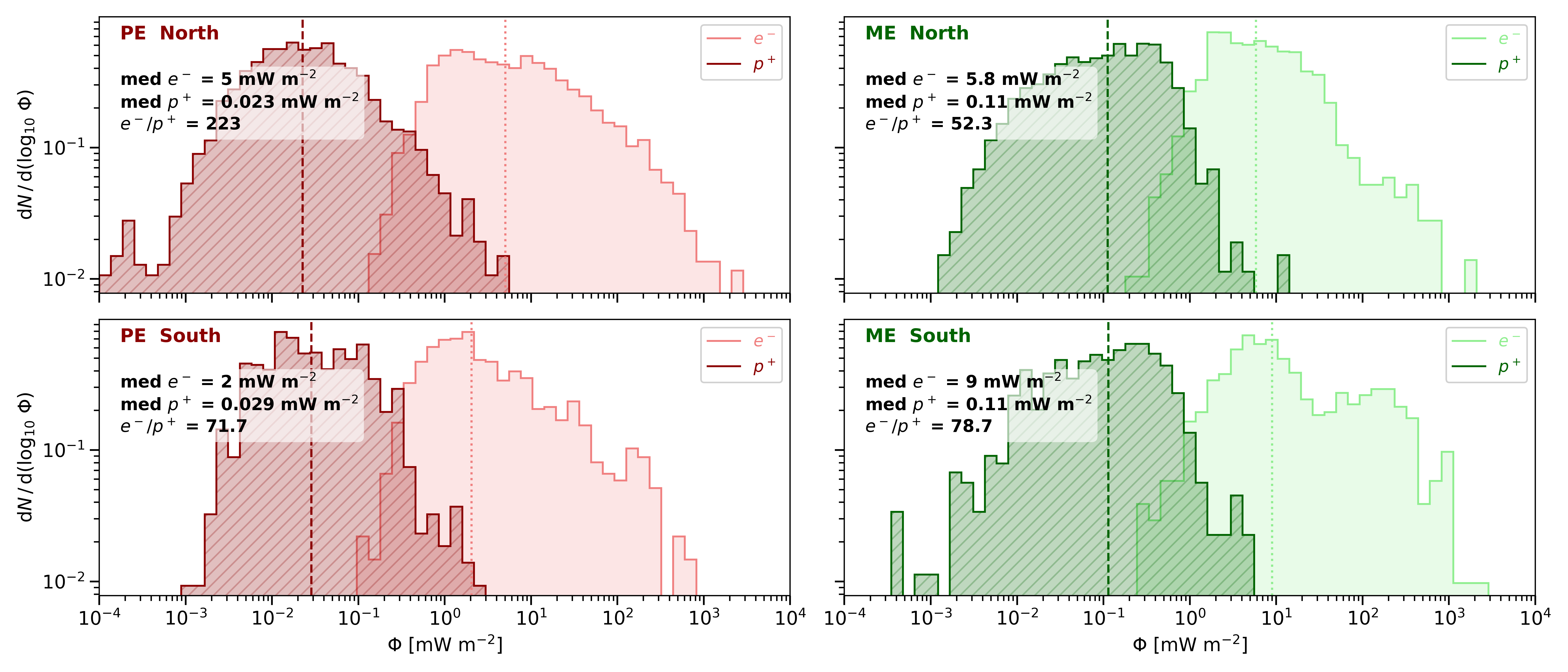}
        \caption{Statistical distributions $\mathrm{d}N/\mathrm{d}\log_{10}\Phi$ of the precipitating energy flux $\Phi_\downarrow$ for electrons (light shading) and protons (dark shading with hatching), in the polar emission (PE, red) and main emission (ME, green) regions, for the northern (top row) and southern (bottom row) hemispheres. Vertical dashed and dotted lines indicate the proton and electron median values, respectively, which are reported in each panel together with their ratio.}
    \label{fig:powerflux}
\end{figure*}

In both sub-regions and in both hemispheres, the proton distributions are systematically shifted toward much lower values than those of the electrons, by a factor of 52 to 223 (Fig.~\ref{fig:powerflux}), corresponding to a global offset of 1.7 to 2.3 orders of magnitude. Since the JEDI-only electron energy fluxes used here are underestimated by $\sim$6\% in median (Sect.~\ref{sec:flux}), the electron side of these ratios is only marginally affected. The proton fluxes also omit the energies below 50\,\keV{}, a contribution that cannot be quantified with JEDI alone, but given median proton mean energies of 144 to 162\,\keV{} (Table~\ref{tab:stats}), this range cannot plausibly carry the factor of 50 or more needed to close the gap between the two species. The upper end of the proton distributions also lies well below that of the electrons ($\sim$1--10\,\mwm{} versus $\sim$10$^3$\,\mwm{}).

In the PE and ME, the proton medians are essentially the same in both hemispheres, 0.023 and 0.029\,\mwm{} in the PE (north and south) and 0.11\,\mwm{} in both hemispheres of the ME, whereas the electron medians differ by a factor of about two in each of these two sub-regions, and in opposite senses, the north dominating in the PE (5.0 against 2.1\,\mwm{}) and the south in the ME (9.0 against 5.9\,\mwm{}). The hemispheric contrast carried by the electrons is therefore not reproduced by the protons, although the sparser proton sampling prevents us from excluding a contrast of comparable amplitude. 

Table~\ref{tab:stats} reports the sample sizes, the dispersion of each distribution, and the mean spectrum energy underlying these medians. The electron and proton spectra have comparable mean energies, differing by only a factor of $\sim$1 to $\sim$1.5 in every sub-region and hemisphere. Since this factor is close to unity, it cannot by itself account for the roughly two order of magnitude gap in energy flux. The energy flux ratios quoted above (52 to 223) are therefore close to the corresponding particle flux ratios, indicating that the disparity between the two species is driven primarily by their particle number flux rather than by a difference in typical particle energy.

\begin{table}[h!]
\centering
\caption{Number of spectra $n$, median precipitating energy flux $\Phi$ with 16th and 84th percentile bounds, and median mean spectrum energy $\langle E \rangle$, for protons and electrons in each auroral sub-region and hemisphere. JEDI data, PJ3--PJ66.}
\footnotesize
\setlength{\tabcolsep}{3pt}
\begin{tabular}{@{}llccccc@{}}
\toprule
 & & & $n$ & $\Phi$ [\mwm{}] & $[\Phi_{16};\Phi_{84}]$ & $\langle E \rangle$ [\keV{}] \\
\midrule
\multirow{4}{*}{Electrons} & \multirow{2}{*}{PE} & N & 3819 & 5.03   & [1.01; 37.1]    & 227 \\
                           &                     & S & 1005 & 2.05   & [0.67; 15.7]    & 238 \\
\cmidrule(lr){2-7}
                           & \multirow{2}{*}{ME} & N & 2123 & 5.85   & [1.88; 24.6]    & 163 \\
                           &                     & S & 760  & 9.00   & [2.78; 127]     & 165 \\
\midrule
\multirow{4}{*}{Protons}   & \multirow{2}{*}{PE} & N & 3481 & 0.0226 & [0.0054; 0.109] & 162 \\
                           &                     & S & 1591 & 0.0286 & [0.0078; 0.122] & 151 \\
\cmidrule(lr){2-7}
                           & \multirow{2}{*}{ME} & N & 1948 & 0.112  & [0.018; 0.430]  & 160 \\
                           &                     & S & 654  & 0.114  & [0.018; 0.418]  & 144 \\
\bottomrule
\end{tabular}
\label{tab:stats}
\end{table}

Energetic protons in the middle magnetosphere exhibit pancake-shaped pitch angle distributions with nearly empty loss cones \citep{shen_energetic_2022}, and their precipitating flux is therefore intrinsically weaker than that of electrons. Electrons are in addition scattered more efficiently by chorus and ECH waves, owing to their much higher cyclotron frequency. Both effects are physical and act independently of the instrumental limitation discussed in Sect.~\ref{sec:instrument}.

\begin{figure*}[h!]
    \centering
    \includegraphics[width=17cm]{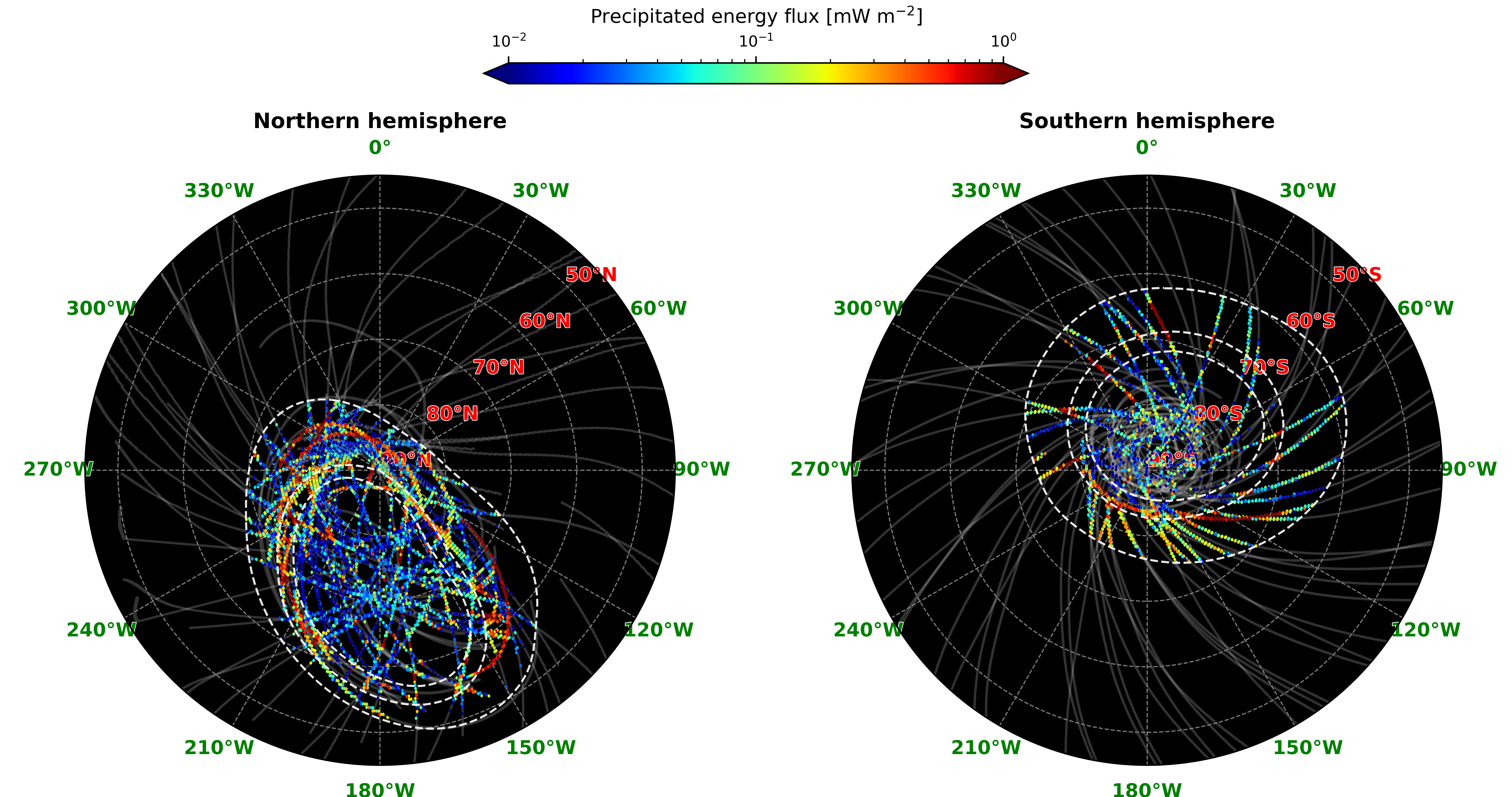}
    \caption{Polar maps of the precipitating proton energy flux $\Phi_\downarrow$ in the northern (left) and southern (right) hemispheres, combining PJ3--PJ66. The color scale is logarithmic. White dashed contours delineate the PE and ME sub-regions.}
    \label{fig:maps}
\end{figure*}

Figure~\ref{fig:maps} presents the polar projection of the precipitating proton energy flux in the northern and southern hemispheres, combining PJ3--PJ66 data. The measurements are broadly distributed across the two sub-regions in both hemispheres, reflecting the overall orbital coverage of Juno over the 66 perijoves. In the northern hemisphere, the proton energy flux is statistically higher in the ME than in the PE, which may reflect more efficient pitch-angle scattering into the loss cone in the middle magnetosphere relative to the polar cap \citep{shen_energetic_2022}, although this interpretation remains tentative given the sparsity of proton measurements. In the southern hemisphere, the PE remains the weakest sub-region.

\subsection{Implications for auroral emissions and the upper atmosphere}

Multiplying the median energy flux of each sub-region by its median area (Sect.~\ref{sec:mapping}) gives the median total precipitated power. For electrons, we obtain $\sim$5.0 and $\sim$1.8\,TW in the PE, $\sim$2.1 and $\sim$3.5\,TW in the ME (north and south, respectively). For protons, the corresponding powers are $\sim$24 and $\sim$26\,GW (PE), $\sim$40 and $\sim$43\,GW (ME). Protons thus account for 0.45--1.9\% of the electron power, that is, less than 2\% in both sub-regions.

For electrons, precipitated power is converted into total UV radiated power with a median efficiency of about 21\% at Saturn \citep{lamy_multispectral_2013}, a value that can serve as an order-of-magnitude analogy for Jupiter given the similar H$_2$-dominated atmospheric composition. Applying this conversion efficiency to the median electron powers estimated here yields UV-radiated powers of order 0.37--1.05\,TW in the PE and 441--735\,GW in the ME. No exact conversion efficiency has been established for the excitation of H$_2$ by proton impact. However, for non-forbidden excitation processes, the average proton-impact conversion is expected to be broadly similar to that of electron impact (G. Gronoff, private communication). Applying the same $\sim$21\% efficiency to the proton precipitated powers above yields first-order UV-radiated power estimates of $\sim$5.0--5.5\,GW in the PE and $\sim$8.4--9.0\,GW in the ME. This estimate excludes the additional contribution from charge-exchange processes, which have no electron-impact equivalent, so a precise estimate would still require a dedicated proton transport model. Such models exist for Jupiter \citep{rego_auroral_1994,houston_jovian_2026}, but were not available to us, and the TransPlanet code used in our previous auroral studies \citep{lilensten_ionization_1989,benmahi_energy_2024,benmahi_auroral_2024} is restricted to electron transport, so that no such refined estimate can be derived within the scope of this paper.

These measurements allow the Voyager and IUE era interpretation to be revisited. Following the first theoretical estimate of a Jovian proton aurora by \citet{heaps_jovian_1975}, \citet{goertz_proton_1980} proposed that proton precipitation could account for a substantial fraction of the auroral emission, an interpretation extrapolated from the trapped population measured by Voyager. The observational tests that followed were indirect. \citet{clarke_doppler_1989} found no red-shifted Lyman-$\alpha$ component in high-resolution IUE spectra, and \citet{waite_role_1992} combined a comparable non-detection with an emission model to conclude that protons contribute less than 10\% of Jupiter's UV auroral emission. None of these studies could measure the precipitating proton energy flux itself, since the models available at the time normalised their results to an assumed input flux \citep{rego_auroral_1994}. The Juno/JEDI measurements reported here supply that missing quantity, and show that the proton UV powers estimated above amount to about 1\% of the electron values, consistent with the IUE upper limit and roughly one order of magnitude below it. The discrepancy with the original Voyager interpretation follows from the nearly empty proton loss cones in the middle magnetosphere \citep{shen_energetic_2022}, which make the precipitating flux far smaller than the trapped flux from which it was extrapolated.

Charge exchange between precipitating protons and atmospheric hydrogen produces fast H atoms whose de-excitation emits a red-shifted Lyman-$\alpha$ photon, with a brightness proportional to the incident proton energy flux \citep{waite_role_1992, gerard_model_2000}. High-resolution HST observations did not clearly detect this component \citep{rego_supersonic_1999,rego_analysis_2001}, consistent with the weak fluxes measured here, and Juno/UVS cannot exploit it due to the localized gain-sag effect observed around Lyman-$\alpha$ prior to Juno's Jupiter orbital insertion (e.g., \citealt{Davis2011, Hue2019a}). The upcoming Juice and Europa Clipper missions will provide the means to measure it directly, using the fluxes reported here, through their respective ultraviolet spectrographs, Juice-UVS \citep{davis_juice_2021} and Europa-UVS \citep{retherford_europa_2024}. Energetic protons also penetrate deeper into the atmosphere than electrons of equal energy, owing to their longer mean free path in H$_2$ \citep{ozak_auroral_2013}, and can ionize and dissociate H$_2$ around the hydrocarbon homopause. The altitude at which each species deposits its energy therefore sets the altitude at which H$_3^+$ is produced. Electron precipitation that drives UV emission also produces H$_2^+$, which reacts within less than a minute with H$_2$ to generate a deep column of H$_3^+$ above the homopause \citep{nichols_dynamic_2025,tiranti_pole--pole_2025}. This density distribution typically closely matches the UV morphology, except for bright and localized UV spots associated with deep precipitating electrons that fall below the homopause \citep{stallard_stability_2016,gerard_concurrent_2018}, and localized regions of very weak precipitation near dusk that are not observed in the UV \citep{nichols_dynamic_2025}. By contrast, the deposition altitude of precipitating protons depends strongly on their energy, peaking near 450\,km for 100\,\keV{} protons and dropping to $\sim$270\,km, in the stratosphere, for 2\,\MeV{} protons \citep{houston_jovian_2026}. The median proton mean energies measured here, 144 to 162\,\keV{}, therefore correspond to deposition close to the hydrocarbon homopause, and only the MeV tail of the spectrum penetrates well below it, where the H$_3^+$ produced is almost instantly destroyed through protonation of hydrocarbons \citep{moore_photochemistry_2026}. H$_3^+$ therefore provides at best limited diagnostic information on proton precipitation.

Given precipitated powers about two orders of magnitude lower than those of electrons, however, their impact on the atmospheric composition and thermal structure remains subordinate. Quantifying it precisely would require coupled proton-electron transport models such as the Trans* codes \citep{lilensten_ionization_1989}, beyond the scope of this paper.

\section{Conclusions}
\label{sec:conclusion}
We present the first statistical comparison of the precipitating energy flux of protons and electrons in the auroral sub-regions of Jupiter, based on Juno/JEDI data from PJ3 to PJ66.
\begin{enumerate}

\item Per magnetic footprint, protons are detected far less often than electrons, owing to a higher detection threshold and the limited efficiency of the mandatory TOF-energy coincidence (peaking at $\sim$0.3 near 90\,\keV{}; \citealp{mauk_juno_2026}), compounded by the intrinsically weaker precipitation of protons from their pancake-shaped pitch angle distributions and weaker wave coupling \citep{shen_energetic_2022}. The comparable final sample sizes of the two species result from the coarser sampling grid adopted for protons.

\item In both sub-regions, the median proton energy flux is lower than that of electrons by a factor of 52 to 223, corresponding to 1.7 to 2.3 orders of magnitude. The total precipitated power of protons amounts to less than 2\% of that of electrons, confirming their secondary role in Jupiter's auroral energy budget.

\item Proton energy flux is broadly distributed across auroral sub-regions, higher in the ME than in the PE.

\end{enumerate}

The energy fluxes reported here provide the quantitative input required to predict the brightness of the Doppler-shifted component in the red wing of the Lyman-$\alpha$ line. This component has so far escaped detection \citep{rego_supersonic_1999,rego_analysis_2001} and lies in a spectral range that Juno/UVS cannot exploit, so that Juice-UVS and Europa-UVS will be the first instruments in a position to search for it.

\begin{acknowledgements}
B. Benmahi and V. Hue acknowledge support from the French government under the France 2030 investment plan, as part of the Initiative d'Excellence d'Aix-Marseille Université, A*MIDEX AMX-22-CPJ-04. French co-authors acknowledge the support of CNES for the Juno and Juice missions. B. Bonfond is a Research Associate of the Fonds de la Recherche Scientifique, FNRS. The authors acknowledge support from the JAFAR project (ANR-25-CE49-6683). T.S. was supported by the STFC Consolidated Grant (ST/W00089X/1).
\end{acknowledgements}
\bibliographystyle{aa}
\bibliography{biblio.bib}

\end{document}